*Review*

# Research on Diamond Open Access in the Long Shadow of Science Policy

Niels Taubert [1*]

[1] Institute for Science Studies, Faculty for Sociology, Bielefeld University, p.o. box 10 01 31, 33501 Bielefeld, Germany; niels.taubert@uni-bielefeld.de

[*] Correspondence: niels.taubert@uni-bielefeld.de; ++49 521 106 4657

**Abstract**

This paper reviews research literature on Diamond Open Access (DOA) journals – sometimes also called Platinum Open Access – that was produced after this journal segment started to become a priority in European research policy around 2020. It contextualizes the current science policy debate, critically examines different understandings of DOA, and reviews studies on the role of such journals in scholarly communication. Most existing research consists of quantitative studies focusing on aspects such as the number of DOA journals, their publication output, the diversity of the landscape in terms of subject areas, languages, publishing entities, indexing in major databases, awareness and perception among scholars, cost analyses, as well as insights into the internal operations of DOA journals. The review shows that research on DOA journals is partly influenced by the science policy discourse in at least two ways: first, through the normativity inherent in that discourse, and second, through the temporality of policy-driven research of practical relevance, which leaves important aspects of the phenomenon understudied. Moreover, research on the DOA journal landscape has implications beyond understanding this particular journal segment, as it also challenges established views of the global system of scholarly communication.

## 1. Introduction

Journals that are free of charge for both authors and readers have existed for a long time. Since the emergence of electronic networks in science – and even before the advent of the internet – scientists have experimented with electronic means to disseminate their research and organize scholarly communication (Björk et al., 2016, p. 1). In addition to electronic preprint servers such as *arXiv*, founded in 1991 (Ginsparg, 1994), the first self-organized electronic journals began operation around 1990. Early examples are *Postmodern Culture* (founded 1990), *Psycologuy* (1990) and *Electronic Transactions on Numerical Analysis* (1993). With the creation of national and regional electronic publishing platforms such as *SciELO* (1998), *J-Stage* (1998), and *Redalyc* (2005), important steps were taken to systematically support such journals with the goal of moving scholarly publishing online and promoting open access (OA) (Packer, 2009, p. 113; Beigel et al., 2024, p. 4). When open access became a topic in science policy, and declarations like the Budapest Open Access Initiative (2002) called for a new generation of journals based on resources other than subscriptions, such journals had already existed for some time (e.g., Rodriguez-Muñoz et al., 2025, p. 35).

Two decades later, actors and initiatives – particularly at the European Union level – adopted the term Diamond Open Access (DOA), which was originally coined by Fuchs and Sandoval (2013), and began actively promoting journals that charge neither authors nor readers. Following the commissioning of a worldwide OA Diamond Journals Study (Bosmann et al., 2021), Science Europe, cOAlition S, OPERAS, and the French National Research Agency launched an Action Plan for Open Access. This plan aimed to develop and expand "a sustainable community-driven Diamond communication ecosystem" (Ancion et al., 2022, p. 3) by fostering "the sharing of infrastructures, standards, policies and practices, [and] specifying suitable quality standards" (ibid., p. 4). At the same time, stakeholders were engaged to "make them aware of their roles in Diamond Open Access" (ibid., p. 5) and to "[d]evelop a framework to ensure that the ownership and governance of Diamond Open Access journal titles and platforms is legally recognized and protected" (ibid., p. 5).

Political support followed from the Council of the European Union, which formulated the goal of an open, equitable, and sustainable scholarly publishing system. In its conclusions, the Council emphasized "that immediate and unrestricted open access should be the norm in publishing research involving public funds, with transparent pricing commensurate with the publication services and where costs are not covered by individual authors or readers" (Council of the European Union, 2023, p. 4). Although this statement allows for publishing models in cooperation with commercial actors, the Council also underlined its preference for models operated by public research institutions, highlighting "[…] the importance of not-for-profit, scholarly open access publishing models that do not charge fees to authors or readers and where authors can publish their work without funding/institutional eligibility criteria" (Council of the European Union, 2023, p. 5).

To many observers, this development came as a surprise, as it set new priorities in the science policy debate on the OA transformation by focusing on a journal segment that has existed for more than a quarter of a century, with some journals having operated for decades. Before this shift – and with few exceptions (Stojanovski et al., 2009) – such journals were only half-heartedly supported in many European countries. Ultimately, the segment was regarded as a niche phenomenon and not central to the OA transformation.

How, then, can this shift in priorities be understood? There is some evidence that the current focus on DOA journals has less to do with their perceived strengths than with growing frustration regarding two alternative pathways toward OA – namely, OA based on article processing charges (APCs) and transformative agreements (TAs) (Dellmann et al., 2022, p. 2). The sources of this frustration are manifold. In terms of costs, the implementation of the APC model has neither resulted in budget relief for libraries nor reduced the overall costs of circulating scientific information. Instead, a steady increase in APC prices has been observed (Jahn & Tullney, 2016, p. 7; Pieper & Broschinski, 2018, p. 11; Borrego, 2023, p. 364), perpetuating the publishing market oligopoly (Butler et al., 2023, p. 786). Moreover, it has become evident that not only the hybrid model (Asai, 2021) but also the gold OA model reproduces global inequities, as publications primarily originate from authors based in high-income countries. Although waivers are theoretically affordable for publishers (Taubert et al., 2021, p. 10), they do not appear to eliminate price barriers for authors in lower-income countries (Druelinger & Ma, 2023, p. 356). Finally, a number of problematic publication practices have emerged that are strongly associated with the APC model, given its close link between the number of articles published and publisher revenue. From a scientific perspective, this creates undesirable consequences when publishers "prioritize profitability over their responsibilities for promoting scholarly communication" (Yoon et al., 2025a, p. 17). Predatory open access journals which publish articles without adequate quality control (Beall, 2012, p. 179) are perhaps the most extreme – though by no means the only – manifestation of this dynamic.

This review paper takes this shift in the science policy discourse as its starting point and focuses on recent studies investigating DOA journals. It asks about the current role of this journal segment, leaving aside more technical aspects that may deserve a separate examination. In doing so, the review aims not only to identify blind spots in the research landscape where further studies are needed but also to criticize the often overly close relationship between science policy discourse and research on DOA, which calls for more reflection. The conclusion may therefore appear somewhat uncomfortable: while research on DOA is undoubtedly necessary to navigate the transformation toward OA, and current studies provide important orientation, such research must become more emancipated from science policy in at least two respects. First, regarding the normativity that accompanies the recent science policy discourse, and second, concerning the time constraints for DOA research, which hinders studying questions that require more complex research designs, including qualitative approaches.

The review is organized as follows. Section 2 explains the selection of materials used in this review. This is followed by a discussion of two aspects of the definition of DOA journals (Section 3): their contested nature and their normative underpinnings. Section 4 discusses studies that analyses the DOA journal landscape, including findings on the number and publication output of DOA journals, followed by an overview of their diversity (Section 5). Section 6 presents findings on the extent to which DOA journals are indexed, followed by an overview of recent studies on how DOA is perceived (Section 7) and cost analyses (Section 8). Section 9 turns to the internal operations of DOA journals and focusses on factors such as reliance on voluntary work, challenges, and journal stability. The discussion (Section 10) selectively highlights results that require further investigation and argues for a stronger emancipation of DOA research from science policy discourse. This is followed by a conclusion.

## 2. Materials and Methods

The sources for this review are publications that report empirical results on the role of DOA journals in the communication system of science and which have been published after DOA became a priority in science policy discourse around 2020. Given that the definition of DOA is contested (see Section 3), publications discussing the meaning of DOA are also included. However, several genres focusing on DOA fall outside the scope of this review. These include empirical studies addressing technical aspects of journal publishing (e.g., digital platforms, identifiers, and standards used by DOA journals), descriptions or developments of DOA infrastructures and related initiatives, opinion papers on Open Access (OA), first-hand accounts from editors, community members, or providers of infrastructure, studies focusing on single institutions, and research focusing on DOA books.

In order to identify relevant studies for this review, three databases were consulted: the Web of Science Core Collection (WoS), Scopus, and OpenAlex. In the first step, all entries were retrieved from the three databases that met the following criteria:

- Publication year between 2020 and 2025
- Publication type: article
- Language: English
- Search term "diamond open access" appears in the title, keyword and/or abstract field
- The article is published in a relevant subject category

For Scopus, relevant subject areas include *Social Sciences*, *Computer Science*, *Arts and Humanities*, and *Business, Management, and Accounting*. For WoS, the relevant categories are *Information Science & Library Science*, *Science and Technology Other Topics*, *Computer Science*, *Communication*, *Social Sciences Other Topics*, *Sociology*, and *History and Philosophy of Science*. In OpenAlex, topics were restricted to *Scientometrics and Bibliometrics Research*, *Research Data Management Practice*, *Academic Publishing and Open Access*, *Scientific Computing and Data Management*, *Cultural, Media and Literary Studies*, and *Publishing and Scholarly Communication*.

With this search strategy, a considerable number of potentially relevant articles were identified (see Table 1). In order to select the final corpus, all abstracts were screened. Studies addressing mainly technical aspects, descriptions or developments of DOA infrastructures, opinion papers, first-hand experiences, and studies focusing on a single institution or DOA books were excluded. After removing duplicates, 59 articles were identified as relevant. This corpus was complemented by a limited number studies of immediate relevance that were cited in the corpus and published as reports.

**Table 1.** Sources of articles relevant for review

| Database | Initial no. of articles | Identified as relevant |
| --- | --- | --- |
| Web of Science | 104 | 28 |
| Scopus | 69 | 38 |
| OpenAlex | 106 | 57 |

## 3. What are DOA journals?

*3.1 Contested Definition*

As is often the case in research on Open Access (OA), some confusion exists. The term Diamond Open Access (DOA) has been used with varying meanings, and there is an ongoing debate about how it should be understood. Moreover, the alternative term Platinum OA (Haschak, 2007) is sometimes used synonymously. Given that current research tends to favor DOA, this term will be used throughout the present paper.

In the literature, there is at least a minimal consensus that DOA refers to publishing models characterized by the absence of costs for both authors and readers (Dellmann et al., 2022, p. 1). This understanding was popularized by the widely cited OA Diamond Journals Study, which defined its object of investigation as "journals that publish without charging authors and readers, in contrast to subscription journals" (Bosman et al., 2021a). A similarly pragmatic definition has been adopted by a number of studies (Hahn et al., 2022, p. 50; Frantsvåg, 2022, p. 1; Taubert et al., 2024, p. 196; Arasteh & Blake, 2024, p. 6; Pilatti et al., 2025, p. 2) and is also used for self-identification by DOA journals (González Tous et al., 2021, p. 2; Gallo & Accogli, 2022, p. I; Biel, 2023, p. 2; Tur-Vines, 2023, p. 10; Fick, 2024, p. 35, Ptakova, 2024, p.127).

However, there are more complex definitions that propose additional criteria beyond the mere absence of payments (Beigel, 2025, p 10). One prominent example is the Operational Diamond OA Criteria for Journals, developed within the context of two Horizon 2020-funded EU projects, DIAMAS and CRAFT-OA. These specify six criteria. In addition to the absence of payments, journals should be scholarly, have a permanent identifier, use an open license, and be open to all authors. These criteria aim to clarify what qualifies as a scholarly journal. Yet one particularly significant criterion that considerably narrows the group of eligible journals is community ownership. This designation is somewhat misleading, since the legal entities that own such journals are in most cases not the scientific communities themselves but rather organizations such as "public or not-for-profit

organization[s] (or parts thereof) whose mission includes performing or promoting research and scholarship" (Armengou et al., 2024).

This ownership criterion addresses an objection raised by Simard et al. (2024, p. 1024), who argue that journals run by commercial publishing houses that temporarily waive APCs should not be considered DOA. It is worth noting that these additional criteria are not justified by research-oriented arguments but rather by reference to practical concerns – particularly fears that large publishing houses might continue exploitative strategies under the label of DOA (Simard et al., 2024, p. 1025). Such an anti-corporatist stance is not new to the discussion but has accompanied the term since its inception. When proposing it, Fuchs and Sandoval (2013) defined Diamond OA as a "not-for-profit" model run by "non-commercial organizations, associations, or networks" that publish material online in digital format, free of charge for both readers and authors, and that do not allow commercial or for-profit reuse. This understanding is reflected in a best practices practice paper on DOA journals (Zendejo & Escalona, 2025, p. 5) as well as in self-descriptions of some journals identifying their model as DOA (Tiwari, 2020, p. 1).

Such definitions may be appropriate in the context of science policy, where the aim is to promote a specific understanding of the OA model. However, with respect to research on OA, definitions should not be guided solely by practical considerations for at least two reasons. First, we have argued elsewhere that it is a general requirement of scientific classifications that they should be complete and mutually exclusive and that such requirements also hold for types of OA (Taubert et al., 2019, p. 7). Otherwise, it becomes necessary to invent additional types of OA for cases that do not meet all criteria that are specified in a given definition, and such cases are manifold: how should we, for example, classify a journal, according to Fuchs and Sandoval's definition, that neither charges authors nor readers but whose title rights are owned by a publisher, while a public library organizes a campaign to fund its operations? It would not be DOA, and in order to classify them, we would need to think of a category like semi-Diamond public–private OA. Or consider a scholarly journal operated by a group of dedicated scholars who, being unaware of or indifferent to copyright issues, fail to provide any license – would this still be considered DOA? According to the Operational Diamond OA Criteria for Journals, the answer would be no, and again a new category would have to be invented – for example, non-legally valid Diamond OA.

Second, and equally important, additional criteria also pose challenges for researchers who aim to consistently identify DOA journals in large data collections. Given the lack of an authoritative and exhaustive database of DOA journals, it is already difficult to verify the absence of payments on both the side of the author and the reader for large journal samples. The inclusion of ownership or other criteria – often missing from databases and not easily ascertainable through web searches – makes the valid identification of DOA journals considerably more difficult.

*3.2 Impregnated with Normativity*

Closely related to definitional issues, it appears difficult for parts of the research on DOA journals to describe their object of investigation without adopting normative language. The sources of normativity are evident: in current science policy discourse, DOA journals are associated with strongly positive attributes and are presented as the desirable publishing model. Two examples shall be given to illustrate this point: the 2nd Diamond Open Access Conference associates seven positive attributes to DOA journals, namely, (1) equity, (2) knowledge as a public good, (3) community-driven initiatives, (4) diversity, (5) transition to diamond, (6) research assessment and recognition, and (7) multi-level cooperation (Saenen et al., 2024). Whether these characteristics are assumed to apply to all journals is not entirely clear; however, there is no doubt that these features are considered as positive and very much welcome. The second example comes from the Diamond Open Access Standard (DOAS), developed within the DIAMAS project. It characterizes DOA as the "ideal, most equitable, end state of scholarly publishing" (Consortium of the DIAMAS Project, 2024, p. 5) but makes clear that even for journals that have entered such normative telos of scholarly publishing, there is room for improvement. This can be achieved by adapting an element that is listed as a "desired" feature. Examples are the adaption of an open science policy that "shows [the journal] is aware of the value of OS and understands what it entails" or the acknowledgement that publications of negative or unexpected results contribute to the advancement of science (Consortium of the DIAMAS Project, 2024, pp. 20-21). Within such contextualization, DOA journals appear to be a universal solution to various types of problems associated with scholarly publishing today.

Such normative attributions spill over into science and are echoed not only in opinion papers (e.g. Andringa et al. 2024, p. 7., characterizing DOA as 'truly inclusive') but also in parts of the research literature. This occurs when certain characteristics are presented as inherent features of DOA journals from the outset, without empirical evidence or further justification. Examples include the characterization of DOA journals as sustainable (e.g., Peruginelli & Faro, 2024, p. 67), equitable (ibid.), or epistemologically diverse (e.g., Stojanovski & Morfardin, 2025, p. 1). The challenges posed by such normative undertones for understanding the operations of DOA journals and for providing evidence-based input for informed decision-making will be discussed later.

## 4. DOA journal landscape

After having discussed two characteristics of the term DOA journals, the following sections present and discuss empirical findings from different types of studies. The first strand of research on DOA consists of studies that analyse the DOA journal landscape that seek to map and characterize the field within specific boundaries. These studies vary in scope: some adopt a global perspective, while others focus on a single country or a particular disciplinary context. The entities being mapped may include journals and, less frequently, organizations such as institutional publishers or service providers. Common topics of investigation include the number of DOA journals, their geographic or disciplinary distribution, as well as their publication output.

Before turning to specific findings, one methodological observation deserves attention. A key challenge for studies that analyse the DOA journal landscape is the lack of an authoritative and comprehensive source of information on DOA journals. The most prominent source for Gold Open Access, the Directory of Open Access Journals (DOAJ), includes an "APC" flag that allows researchers to identify journals that do not charge authors. However, this data source is far from exhaustive. In response, many studies adopt a similar strategy: they combine and cross-check multiple data sources.

For example, Bosman et al. (2021a, p. 26) use the ISSN Gold OA 4.0 list (Bruns et al., 2020), which integrates data from DOAJ, ROAD, PMC, and OpenAPC. Similar approaches are found at the national level. A Swiss study combines DOAJ, Swisscovery, the SCImago Journal Ranking list, EBSCO Academic Search Ultimate, Web of Science, and ROAD (Hahn et al., 2022, p. 10); an Italian study draws on the Electronic Journals Library (EZB), ROAD, OpenAIRE Graph, AVENUR List of Classified Journals, and SCImago (Angioni et al., forthcoming, p. 8); a German study again uses DOAJ, ROAD, PMC, and OpenAPC (Taubert et al., 2024, p. 198); a Canadian study combines OpenAlex, the CRKN Open Access Journal Lists, Érudit as well as data from the Public Knowledge Project (van Bellen & Céspedes, 2025, p. 98); and a study of South American and Caribbean countries uses SciELO and Redalyc (Beigel et al., 2024). These examples illustrate that substantial effort is still required to create databases that are suitable for the systematic analysis of DOA journals.

*4.1 Number of Journals*

The number of DOA journals reported in the literature is impressive, regardless of the entities examined. At the global level, Bosman et al. (2021a, p. 27) estimate the number of DOA journals to range between a lower bound of 17,000 and an upper bound of 29,000 – a magnitude confirmed by estimates from the Korean Council of Science Editors (2022, p. 123). Of the 11,064 journals indexed in DOAJ, approximately 45% are published in Europe, 25% in Latin America, 16% in Asia, and nearly 6% in the United States and Canada (Bosman et al., 2021a, p. 32). Based on DOAJ data, a large increase in the number of Diamond OA journals can be observed between 2015 and 2020 (Simard et al., 2022, p. 3).

At the country level, 186 DOA journals have been identified for Switzerland (Hahn et al., 2022, p. 11), 205 for Pakistan (Raza et al., 2022, p. 4), 298 for Germany (Taubert et al., 2024, p. 201), 462 for Italy (Peruginelli & Faro, 2025, p. 70), and 571 for Canada (van Bellen & Céspedes, 2025, p. 104). A study of 1,720 journals hosted by the South American and Caribbean publishing platforms SciELO and Redalyc found that the vast majority – 1,538 journals – are DOA, charging no fees to authors (Beigel et al., 2024, p. 18). The largest numbers of journals on the two platforms are reported for Brazil (506), Colombia (291), Mexico (283), Argentina (167), and Chile (144), reflecting a long and strong tradition of DOA publishing in those countries (Beigel et al., 2024, p. 10). However, these absolute numbers cannot be directly compared, as both the data sources and the criteria used to assign journals to countries differ between studies.

Khanna et al. (2022) take a different approach and focus on journals using Open Journal Systems (OJS), an editorial management and hosting platform developed by the Public Knowledge Project (PKP). The platform is frequently used by journals operating independently of major publishers. The number of active journals in 2020 was as high as 25,671 (Khanna et al., 2022, p. 914), distributed across 136 countries. The largest numbers are found in Indonesia (11,535 journals) and Brazil (2,653). More than 81% of OJS-based journals are located in middle- or low-income countries (Khanna et al., 2022, p. 920).

Beyond absolute numbers, several studies also examine the share of DOA journals among all Gold OA journals. At the global level, estimates range from 66% (Crawford, 2024, p. 6) to 73% (Bosman et al., 2021, p. 26). Evidence regarding trends over time is less straightforward. A study based on DOAJ data found a steady decrease in the share of DOA journals between 1986 and 2020 (Druelinger & Ma, 2023, p. 352). For the subset of journals indexed in the Web of Science, only 3% of journals with a Journal Impact Factor (JIF) are DOA. This share is, however, growing more rapidly than the overall number of academic journals in that subset (Pearce, 2022, p. 212). These seemingly contradictory trends may reflect the selective inclusion of journals in the Journal Citation Reports or the faster growth of APC-based Gold OA journals. At the national level, shares of DOA also vary: in Canada, the share is 61% (van Bellen & Céspedes, 2025, p. 104), while in Pakistan it stands at 69% (Raza et al., 2025, p. 104), roughly aligning with the global average.

Some studies adopt a disciplinary perspective, examining DOA journals in relation to other types such as APC-based Gold OA, hybrid, or subscription journals. For engineering, 757 active OA journals were identified, of which 504 (67%) charged APCs and 253 (33%) were DOA – representing 8.4% of all engineering journals indexed in Scopus (Pilatti et al., 2025, p. 3). A comparison across quartiles based on CiteScore metrics revealed that DOA journals are less frequent in higher quartiles and overrepresented in lower ones (Pilatti et al., 2025, p. 5). Further analysis of top-quartile journals showed that article output is strongly dominated by APC-based mega-journals published by the scholarly association IEEE (Pilatti et al., 2025, p. 6). However, due to the limited coverage of DOA journals in proprietary databases, their actual share in this disciplinary segment may be underestimated.

*4.2 Publication output of Diamond OA journals*

The analysis of the number of journals provides insight into the scale of ongoing efforts to offer fee-free publishing venues, but it does not indicate the overall relevance of DOA within the broader system of scholarly communication. Publication output can range from only a few articles per year to several tens of thousands in the case of mega-journals.

When article volume is related to journal numbers, studies consistently find that although the majority of Gold OA journals do not charge fees, most articles are published in APC-based journals – a phenomenon Crawford (2024, p. 6) describes as the "articles-vs.-journals split." While 63.8% of Gold OA journals listed in DOAJ do not charge a fee, 70.5% of articles are published in APC-based journals. According to a DOAJ-based study, DOA journals accounted for the majority of article output in Gold OA journals from 1986 onwards. Since 2008, however, this relationship has shifted: APC-based Gold OA journals now publish more than half of all articles in the database, rising to 68.7% in 2020 (Druelinger & Ma, 2023, p. 351). When interpreting these results, it must be considered that DOAJ coverage is not exhaustive and may be biased in favor of APC-based journals.

While DOA journals are numerous, their publication output is comparatively modest. For DOA journals indexed in DOAJ, the annual average number of articles was 34 for 2017–2019, compared with 55 for APC-based journals. Large journals with annual outputs exceeding 500 articles are virtually absent (Bosman et al., 2021a, p. 36). Similar findings are reported elsewhere: in Switzerland, the median output was 15 articles in 2021, based on a survey of editors (Hahn et al., 2022, p. 20). For Italy, Angioni et al. (forthcoming) also report small-scale publishing activity, with most journals publishing between 10 and 24 articles per year between 2000 and 2024, albeit with a gradual tendency toward expansion. For German DOA journals, the median output in 2021 was 16 articles per year, compared with 72 for journals indexed in the Web of Science. Large-scale journals with outputs above 500 articles annually are absent from the German DOA landscape, whereas more than 5% of journals indexed in the Web of Science exceed this threshold (Taubert et al., 2024, p. 206). These findings align with a study by Björk et al. (2016), who examined 250 "indie" journals – initiatives launched by scientists without the backing of publishers or learned societies before 2002. The study reports an increase in median article output from 12 in 2003 to 18 in 2014 (Björk et al., 2016, p. 7). Thus, the majority of DOA journals can be considered small- to mid-sized publishing ventures.

Moreover, there is evidence that journal size correlates with the presence or absence of fees, and that this, in turn, correlates with the size of the publishing entity (in terms of the number of journals published). A study focusing on Europe finds that DOA journals are more often published by small and mid-sized publishers, while APC journals are much more frequently published by larger publishing companies (Laakso & Multas, 2023, p. 452). This result is confirmed by a global study, which also highlights strong disciplinary differences. In the arts and humanities, 77% of articles in journals indexed in DOAJ are published in DOA journals of small publishers, whereas the share is as low as 8% for large publishing companies. In the social sciences, the difference is similarly pronounced, with 65% of articles published in DOA journals of small publishers compared to 16% in those of large publishing companies (van Bellen & Cespedes, 2025, p. 11).

Some studies also estimate the share of DOA publications in the overall scholarly output. Compared with the total global journal output – including hybrid and subscription titles – DOA journals account for an estimated 356,000 articles per year for 2017–2019, corresponding to 8–9% of all scholarly articles (Bosman et al., 2021a, p. 30). Findings at the country level are in line with this result: in Italy, 9.4% of national research output appears in DOA journals (Angioni et al., forthcoming, p. 10), while Frantsvåg (2022, p. 4) reports an increase in the share of Diamond OA publications in another national context from 5% in 2017 to 8% in 2020.

In summary, studies that analyse the DOA journal landscape provide an essential empirical foundation for understanding the scope and evolution of Diamond Open Access publishing. Despite methodological challenges in constructing reliable datasets, these studies consistently demonstrate that DOA represents a substantial segment of scholarly publishing. While the number of DOA journals is large and continues to grow across regions, their individual publication volumes remain comparatively modest (Armengou et al., 2023, pp. 25, 53), with some countries and regions being exceptions.

## 5. Diversity of Diamond Open Access Journals

Building on the quantitative overview presented in the previous section, this section examines the diversity of DOA journals in more detail, focusing on variations across disciplines, languages, and publishing entities.

*5.1 Distribution among subject areas*

A first dimension of diversity concerns the subject areas, disciplines, or scientific fields covered by DOA journals. On a global level, notable differences can be observed. The main concentration lies in the social sciences and the arts and humanities, which together account for 61% of DOA journals. Twenty-two percent publish research in the sciences and 17% in the humanities (Bosman et al., 2021a, p. 34). These findings confirm earlier results by Björk et al. (2016, p. 13). More disaggregated data at the regional and national levels, however, reveal a more nuanced picture. In many European countries, the distribution mirrors the global pattern. For example, in Switzerland, DOA journals are strongly concentrated in the social sciences (44.6%), arts and humanities (21.0%), and life sciences and biomedicine (21.0%) (Hahn et al., 2022). A similar pattern is observed in Germany, where 36.9% of DOA journals publish in the social sciences and 35.6% in the arts and humanities (Taubert et al., 2024, p. 203). Recent studies from Italy (Angioni et al., forthcoming, p. 9), Canada (van Bellen & Céspedes, 2025, p. 104), and Norway (Frantsvåg, 2022, p. 4) confirm comparable disciplinary profiles, often summarized as a "predominance of SSH disciplines" (van Bellen & Céspedes, 2025, p. 104).

However, three studies provide empirical evidence that this global distribution is largely shaped by European DOA journals, whereas other regions exhibit distinct disciplinary profiles. A study of 25,671 journals using the Open Journal Systems (OJS) platform – of which 84.2% are DOA – found that the social sciences account for 45.9% of journals, while science, technology, engineering, and mathematics (STEM) fields represent a substantial 40.3%. A re-analysis of this dataset by Bosman et al. (2021), cited in Yoon et al. (2025b, p. 56), identified clear regional differences: whereas the social sciences dominate in Europe, science journals are more prevalent in Africa, Asia, and the Middle East, where their share ranges between 23.1% and 30.4%. Beigel et al. (2024, p. 14) examined Latin American and Caribbean journals hosted on the SciELO and Redalyc platforms. The vast majority (89.4%) of the 1,720 journals are DOA (see also Segovia et al 2024, p. 4). Across the region, social sciences account for 33% of journals, medical and health sciences for 19%, humanities for 12%, and natural sciences for 10%. Analyses on the country level further reveal distinct national profiles: social sciences dominate in Uruguay, Ecuador, Puerto Rico, Brazil, Bolivia, Mexico, and Peru, whereas medical and health sciences are particularly well represented in Cuba, Uruguay, Bolivia, Costa Rica, and Peru. In contrast, Argentina, Bolivia, and Brazil show relatively high shares of journals in the natural sciences.

These differences in the disciplinary composition of DOA journal landscapes underscore the importance of country-level analyses. While DOA journals in many regions focus primarily on the social sciences and humanities, significant regional and national variations exist –particularly in Latin America, the Caribbean, and parts of Asia. In order to better understand the adaptability of DOA journals to diverse scholarly contexts, it is essential to identify the driving factors behind both their disciplinary orientations and their country-specific characteristics.

*5.2 Language*

A second dimension of diversity concerns the languages accepted for publication. Empirical evidence consistently indicates that DOA journals display greater linguistic diversity than APC-based journals. A global study comparing journals indexed in DOAJ found that 85.5% of APC-based journals and 62.2% of DOA journals accept only one language. Two languages are accepted by 9.7% of APC-based journals and 23.1% of DOA journals, while three languages are accepted by 3.6% and 9.2%, respectively. English is the most common publication language, used by 70.5% of DOA journals compared to 90.7% of APC-based journals. Among DOA journals, 25.7% accept Spanish, 17.4% Portuguese, 9.1% Indonesian, and 9.1% French. For APC-based journals, only Indonesian stands out, with 9.6% of journals publishing in this language (Bosman et al., 2021, p. 41).

These findings are supported by other studies. A comparative analysis of DOA journals in English-speaking and non-English-speaking countries found that 19.8% publish exclusively in English, 55.9% in English and other languages, and 24.3% in non-English languages (Yoon et al., 2025b, p. 60). Similar patterns appear among small publishers – often responsible for DOA journals – where multilingualism is common (Laakso & Multas, 2023, p. 453). Studies of OJS-hosted journals also confirm this diversity: 60 languages are represented, and nearly half (48%) of the journals allow publication in more than one language (Khanna et al., 2022, p. 920). Supporting evidence for large language diversity in DOA journals is provided by a study of SciELO and Redalyc journals which shows that, among articles published between 1995 and 2018, 43.7% were written in Spanish, 32.1% in Portuguese, and only 23.9% in English (Beigel et al., 2024, p. 26). Multilingualism also characterizes DOA publishing in certain European contexts. In Switzerland, 36.6% of DOA journals publish exclusively in English, while 44.1% accept more than one language (Hahn et al., 2022, p. 14). In Norway, Nordic languages play an important role in most scholarly fields, with the notable exception of the natural sciences and engineering (Frantsvåg, 2022, p. 9). Overall, these findings highlight the integral role of linguistic diversity in shaping the global DOA landscape.

To summarize, the analysis of linguistic diversity suggests that the global communication system of science is less homogeneous than often assumed by bibliometric studies. Divisions are evident not only between language regions but also within multilingual countries. Latin America and the Caribbean provide a clear example, as scholarly publishing frequently occurs in languages other than English. One may conclude that these countries "have established their own scholarly community centered around their languages" (Yoon et al., 2025b, p. 63).

*5.3 Publishing entity*

A third dimension of diversity concerns the institutions that serve as publishers of DOA journals. The sources of information used to identify these publishing entities are likewise diverse. Some studies draw on existing data collections, such as those from publishing platforms like SciELO and Redalyc (Beigel et al., 2024), while others rely on open access evidence sources such as the DOAJ (Bosman et al., 2021, p. 35). Still others collect data independently, for instance, through surveys with journal editors or service providers (Stojanovski & Mofardin, 2025), or by gathering information directly from DOA journal websites (Taubert et al., 2024).

According to an analysis based on the DOAJ, 71.4% of DOA journals are published by universities, 13.7% by OA publishers, 10.1% by societies and governmental organizations, and 3.6% by commercial publishers at the global level (Bosman et al., 2024, p. 35). For Switzerland, 47.3% of DOA journals are published by higher education institutions, followed by 19.9% by academic societies, and a notable 11.3% by for-profit publishers (Hahn et al., 2022, p. 13). Parts of these findings align correspond to the situation in Germany, where the majority of DOA journals are published by research institutions (57.7%) and 10.4% by learned societies. However, the share of DOA journals published by commercial publishers is even higher than in Switzerland, at 17.1%, which can partly be attributed to the strong adoption of the Subscribe-to-Open model in Germany. Moreover, it is worth noting that 5.0% of German DOA journals are published by individuals.

In Italy, a major share (47.7%) of DOA journals is published by institutions, with an additional 26.2% also attributed to institutions (Angioni et al., forthcoming, p. 13). The proportion of journals published by commercial companies is 4.2%, which is close to the global average. The strong role of public organizations also becomes apparent when the focus shifts from journals to providers of infrastructure. A survey of 77 respondents reports that most institutional publishers in Croatia are public organizations, followed by private, not-for-profit organizations (Stojanovski & Mofardin, 2025, p. 5).

In Latin American and Caribbean countries, the situation is similar: journals hosted on publishing platforms such as SciELO and Redalyc are predominantly published by public organizations. Universities account for 66% of journals on these platforms, independent research centers for 20%, and governmental agencies for 7%. There is also a notable share of 8.5% of journals published by commercial publishers. However, it is not clear whether all of them are DOA, as 89.4% of journals on the two platforms report that they do not charge publication fees (Beigel et al., 2024, p. 17).

In summary, the diversity of DOA journals manifests across multiple, interrelated dimensions – disciplinary, linguistic, and institutional. While the social sciences and humanities continue to dominate the global DOA landscape, significant regional variations reveal a more complex and pluralistic role for DOA journals across countries and world regions. Likewise, the linguistic practices of DOA journals underscore their cultural embeddedness: multilingual publication and the use of local or regional languages remain common, challenging the dominance of English and enabling the circulation of knowledge in diverse linguistic contexts. At the institutional level, DOA publishing is primarily driven by universities, research institutions, and other public or non-profit actors, although national differences – such as the stronger presence of commercial publishers in some contexts – may reflect distinct policy frameworks and funding mechanisms. Together, these patterns highlight that DOA is not a uniform model but a rich, multifaceted segment of scholarly communication with locally shaped specificities.

## 6. Indexing

A number of studies have examined the coverage of DOA journals in multidisciplinary bibliographic databases, regional indexes, library discovery services, and OA evidence systems. Although their findings generally indicate low levels of coverage, the degree of clarity and comparability among these studies varies for at least two reasons: the methods applied for data collection and the specific data sources used, particularly when relying on existing large-scale databases.

Regarding data collection, some studies are based on surveys and self-reported inclusion in indexes as reported by editors, while others use direct evidence such as ISSN matching. For instance, based on survey responses from editors of 1,359 journals, Bosman et al. (2021) found that more than 1,000 DOA journals were indexed in the DOAJ, and over 900 were listed in Google Scholar. Approximately 500 DOA journals were indexed in Scopus and more than 300 in the Web of Science (WoS) (Bosman et al., 2021, p. 45; Korean Council of Science Editors, 2022, p. 132). However, the validity of these results may be questioned. A key limitation of the survey method is that editors are likely to better recall indexing in well-known databases, while inclusion in lesser-known indexes or service systems may be forgotten or entirely unknown to respondents.

A study that makes use of a large database – the DOAJ – calculated that 87% of DOA journals are covered by OpenAlex. In contrast, coverage in proprietary databases is far more selective: only 22% of DOA journals are represented in WoS, and 28% in Scopus (Simard et al., 2023a). A subsequent study by the same authors found an improvement in coverage but also highlighted significant disciplinary differences. Specifically, 53.2% of DOA journals in biomedical research, 54.9% in natural sciences and engineering, and 68.2% in the social sciences and humanities are not indexed in either WoS or Scopus. The study also shows that the share of indexed DOA journals decreases with lower national income levels – an effect that can be observed across all scientific domains (Simard et al., 2025, pp. 741, 743).

When interpreting these findings, it is important to note that the databases that are used to identify DOA journals are themselves selective, meaning that results may vary when alternative sources are applied. The estimate that roughly two-thirds of DOA journals worldwide are not indexed in DOAJ (Becerril et al. 2021, p.11) is supported by evidence from Canada, where only 29% of DOA journals are listed in this directory (van Bellen & Céspedes, 2025, p. 10). When the sample is restricted to journals using a common publishing infrastructure, the share of non-indexed titles is confirmed. For example, in a study focusing on journals using OJS, 30.9% of the journals were covered by DOAJ and 29.4% by ROAD. Coverage by proprietary databases was even lower, with a share of just 1.1% included in the WoS Citation Report service and 3.9% in Scopus. Even OpenAlex, known for its extensive coverage, indexed only 13.2% of OJS journals (Khanna, 2022, p. 923). However, this figure may have increased, given OpenAlex's current rapid development. A discipline-specific study further supports these patterns. In the field of library and information science – a discipline that is particularly sensitive to issues of indexing – only 3.3% of DOA journals are indexed in WoS, compared to a substantially higher 24.7% of Gold Open Access journals (Yoon et al., 2025a, p. 21).

In sum, the indexing coverage of DOA journals remains low, particularly in proprietary databases. This limited representation poses a significant disadvantage – not only by reducing the visibility and discoverability of these journals but also because it restricts their participation in widely used citation metrics. In disciplines where journal selection is strongly influenced by indicators such as the JIF, the absence of DOA journals from major indexing systems poses a serious obstacle to their development and recognition.

## 7. Awareness and Perception of DOA journals

Partly related to the coverage of indexing are studies that explore their awareness and perceptions of DOA journals, as in some disciplines in the natural and medical sciences, indexing in major databases is closely connected to the journal's reputation. This line of research derives from the broader and long-standing interest in scientists' attitudes toward OA in general (Schroter & Tite, 2006; Creaser et al., 2010; Serrano-Vicente et al., 2016). For the more specific topic of the awareness and perception of DOA journals, one can state that the numbers of participants in current surveys tend to be small.

A survey study from Switzerland reports that 64% of 123 respondents perceive the publication process in DOA journals as quite similar or only slightly different compared to non-DOA journals (Hahn et al., 2022, p. 29). The scientific impact of DOA journals is generally regarded as lower than that of Gold, Hybrid, or closed-access journals – with the exception of the Arts (Hahn et al., 2022, p. 31). High awareness of DOA journals is also reported in a small study (n = 109) conducted among scholars in Iran. Seventy-eight percent of respondents indicated awareness of DOA, primarily associating it (correctly) with free accessibility for readers and the absence of APCs (Subaveerapandiyan et al., 2025, p. 8). The main perceived benefits are that DOA "fosters innovation and advancement in research fields" and "encourages collaborations among researchers from different countries, making scholarly communication more equitable globally" (Subaveerapandiyan et al., 2025, p. 9). The most prominent concerns are the rigor of editorial processes and "uncertainty about long-term sustainability" (Subaveerapandiyan et al., 2025, p. 13). Perceptions of gold and DOA journals were also investigated in a survey among researchers who had published in high-impact journals of a major OA publisher. Again, the sample size was relatively small, with 152 completed responses. Attitudes toward Gold OA were generally positive, except in relation to the transparency of APC pricing and the extent of institutional financial support for APCs. Perceptions of DOA journals were even more favorable, as these journals were seen as ensuring equitable access for both authors and readers while removing financial barriers. However, respondents expressed mixed views on the sustainability of DOA journals and the extent of institutional support they receive (Kumari & Subaveerapandiyan, 2025, p. 64).

Empirical studies also adopt a disciplinary perspective. A small study of researchers in urology (n = 82) found that 54% of respondents expressed a "positive" to "strongly positive" attitude toward OA. Among the 74% who had experience publishing in OA journals, most reported being either "satisfied" or "completely satisfied" with Gold (65%), Diamond (86%), and hybrid (83%) OA journals (Guennoun et al., 2024, p. e1273).

These results of this small survey of studies indicate that DOA journals are established within their particular context. To a large extent, they are perceived as positive and accepted as a channel for communicating research. Moreover, none of the studies provided any empirical results that point to perceptions of the model that could be interpreted as representing an obstacle.

## 8. Cost analysis

One obvious question about DOA journals relates to the origin of the resources that are necessary in order to sustain the operation. If DOA journals neither involve subscription fees nor APC, where do the resources come from? In the literature, monetary resources and unpaid labor (sometimes also called in-kind contributions) are mentioned as possible sources. The first type of resources is discussed here, the second type and possible effects in the subsequent section.

Empirical evidence for the assumption that the adaption of DOA journals results in a decrease of costs for the dissemination of research (e.g. Mrša, 2022, p.2) can be found in a worldwide survey with 1,370 editors of DOA journals. Accordingly, the overall costs for running these journals are low. For over 60% of the 965 journals for which data are available, annual costs are below $/€10,000 including in-kind contributions. (Bosman et al., 2021, p. 110). Twenty-six percent even reported costs lower than $/€ 1,000. When relating the costs to the article output, median costs of $/€208 per article were calculated across all journals. Given that costs vary by journal size and decreases for journals with larger article outputs, the median costs range from $/€556 for journals publishing five to nine articles per year to $/€48 for journals publishing more than 100 articles per year. (Bosman et al., 2021, p. 112). DOA journals around the globe are also supported by paid staff, but again, the scale is small: 53% of the journals report staff of less than 1 Full Time Equivalent (FTE), 28% from 1-2 FTE, 16% 3.5% FTE 2% 6-9FTE and 1% of the journals 10-20 FTE. (Bosman et al., 2021: 114). However, these results point to possible inconsistencies of the data: a share of 60% of the journals with a budget below $/€10,000 might not align to a share of 47% of the DOA journals that report a size of paid staff with more than one or (up to 20) full time equivalents – at least in Western countries with high wages a high per capita income. For Switzerland, costs are reported by a small group of 28 journal editors and the median cost is with CHF 15,000 higher than reported for the global study. However, the variation is large and within the small journal group, some of them – in particular in life sciences & Biomedicine and Physical Sciences – report costs of several thousand CHF (Hahn et al., 2022, p. 39).

The DIAMAS IPSP study in Europe follows a different approach as the focus is not on journals but on institutional publishers and service providers. Such organizations are frequently but not always or exclusively involved in DOA publishing. They often publish a variety of scholarly outlets and in most cases more than one DOA journal. Roughly three quarters of the 685 respondents were institutional publishers that were responsible for at least one journal while the remaining quarter identified itself as a service provider. The majority of the IPSP are small in size, with around 50% having one to five full-time equivalents, and 25% having no paid staff (Atasteh & Blake, 2024, p. 21). When considering regional distribution, it turns out that the share of IPSP that operate without paid staff is considerably higher in Southern and Eastern Europe than in Western and Northern Europe (Armengou et al., 2023, p. 47).

For IPSP, DIAMAS survey study was analysed separately. Again, the scale of resources that are reported by 42 of the 251 Croatian IPSP tends to be small: 9 of them report a budget between 1-10 thousand €, 13 between 11-50 thousand € and another 7 IPSP between 5-100 thousand €. However, one remarkable point is that nearly a quarter of the 42 IPSP did not know the annual budget or did not wish to disclose it even though the results were reported anonymously (Stojanovski & Mofardin, 2025, p. 6).

## 9. 'Voluntary' Work, Challenges, and the Stability of DOA journals

Compared to the relevance attributed to DOA journals, relatively little is known about the specific characteristics of their internal operations. Some studies highlight a strong reliance on unpaid work or volunteer labor (Armengou et al., 2023, p. 46; Hahn et al., 2022, p. 8). In the global survey of 1,379 editors of DOA journals cited above, 60% of the respondents indicated the involvement of volunteers in journal operations, whereas 40% reported not involving them. Among the journals that do involve volunteers, a large majority of 86% reported either a high or medium reliance on voluntary labor (Bosman et al., 2021a, p. 115). In most cases, voluntary and paid work are combined. The main tasks performed by volunteers include editing, proofreading, copy editing, design, and typesetting (Bosman et al., 2021a, p. 117). The study also asked about sources of financial support. Interestingly, the three most important types of organizations identified were organizations performing research, national funding or government agencies, and publishers (Bosman et al., 2021a, p. 117).

Regarding the extent of monetization and voluntary work, a study of 260 DOA journals arrives at more nuanced results by distinguishing between 26 'publication acts' and asking whether money is used. While the degree of monetized task completion is low for publication acts contributing to certification, legal aspects of publishing monetization are higher in publication acts related to material production and technical issues (Dufour et al., 2023, p. 41).

However, evidence from other world regions suggests that the global landscape is not homogeneous with respect to the reliance on voluntary work. Journals hosted on SciELO and Redalyc are not predominantly voluntary enterprises: "There is considerable consensus within Latin America about the need for national and institutional policies to sustain the management and regularity of the journals, but also a strong consensus that the solution is not in the APC model" (Beigel et al., 2024, p. 19). This statement suggests that funding structures for DOA journals in this region tend to be more favorable than in other parts of the world, where the maintenance of DOA journals often "relies on a heterogeneous and often unstable combination of resources, including various forms of institutional funding and grants from projects that are typically time-limited" (Caravale, 2025, p. 437).

The stability of DOA journals is questioned in the literature, particularly with respect to the difficulties of obtaining permanent funding, and sometimes based on empirical evidence. One example is a study of 250 journals launched by scientists without the support of publishers or learned societies before 2002 (Björk et al., 2016). Within this group, a high degree of volatility was observed: by 2014, 100 journals had ceased operation and another 23 had disappeared. Of the remaining 127 journals (50.8% of the original sample), a majority of 115 still published articles in open access, although 8% of them had introduced APCs. A small share of 9% had become subscription journals by 2014 (Björk et al., 2016, p. 6).

The follow-up question of whether the share of ceased journals is higher among DOA journals than among scientific journals overall has been addressed by other studies. Yoon et al. (2024) compared formal characteristics of journals with low and high sustainability. Based on a multivariate regression analysis, they identified seven factors that influence sustainability: region and official language of the publication country, discipline, ownership, financial status, allowance of text and data mining of full-text articles, and open sharing of research data (Yoon et al., 2024, p. 4). Journals published in Western and Eastern Europe, as well as in Latin America, were found to be less sustainable than those in Canada. Journals in the humanities and social sciences, on the other hand, were more sustainable than those in the medical sciences. Moreover, government-owned journals were more sustainable than those published by universities or for-profit publishers (Yoon et al., 2024, p. 5). A second recent study compared different segments of the journal landscape, including DOA journals, and identified journal age as the main predictor of journal stability. DOA journals appeared to be the most vulnerable group, as they have been overrepresented among ceased journals since the early 2000s (van Bellen & Céspedes, 2025, p. 103).

To summarize, the literature provides evidence that DOA journals often rely on unpaid work and tend to be less stable. However, there is limited evidence regarding whether and how unpaid work may relate to journal stability. One such indication is provided by Hahn et al., who quote an editor describing their journal as "run by exploitation," explaining: "I mean, specifically my two colleagues have their PhD dissertations to write. They're doing the bulk of the work. And I think that's sometimes difficult to balance" (quoted in Hahn et al., 2022, p. 18). A second indication comes from a study involving eighteen editors of DOA journals in Aotearoa/New Zealand, which highlights signs of underfunding, including a lack of recognition and insufficient financial, copyediting, technical, training, marketing, and overall support (Hayes & Murdoch, 2025, pp. 8–13). Finally, in the German DOA study, we identified possible mechanisms through which reliance on unpaid contributions may hamper the long-term stability of DOA journals. First, the unpaid nature of contributions limits editors' ability to enforce task completion and to influence how tasks are performed. Second, in cases of resource shortages, workload is often shifted onto already committed team members. Finally, such shifts may result in situations in which particular team members become indispensable (Taubert et al., 2024, pp. 215–219). Given that these findings are based on a small interview sample and a specific national context, further research is needed to better understand the causes of instability among DOA journals.

## 10 Discussion

The review of existing strands of research shows that significant efforts have been undertaken to study DOA journals. Such research is of immediate relevance because it provides essential information for shaping the future (D)OA landscape. The current focus on large quantitative studies that analyse the DOA landscape and survey studies needs to be maintained – not only by extending it to countries, disciplines, and publication infrastructures that have not yet been addressed but also by revisiting parts of the landscape that have already been examined given the fast evolution of DOA. As infrastructural support for DOA journals increases, data availability, previously perceived as a major limitation (Frantsvåg et al. 2022), is expected to improve. Important building blocks include advancements in open bibliometric data (particularly OpenAlex), the expansion of CRIS implementations, and the emergence of new DOA evidence systems such as the EZB, the Diamond Discovery Hub (Bargheer et al., 2024), and journal registries developed by national initiatives (e.g., SeDOA).

However, the review of the literature also reveals a number of open questions and blind spots, making it necessary to recalibrate future research efforts. Above all, a stronger emancipation from the DOA science policy discourse is needed in at least two respects. First, the policy discourse is strongly normative, frequently characterizing this type of publishing as "high-quality," "transparent," "trustworthy," and "equitable" (e.g. Council of the European Union 2023). Too often, such framings are echoed in academic research on DOA. A more emancipated perspective would critically reflect on these normative assumptions; however, this does not yet happen consistently. Various normative biases persist:

- The very definition of the object of study may implicitly exclude the involvement of commercial actors.
- Data collection may focus exclusively on monetary costs while ignoring non-monetary contributions, thereby suggesting that DOA journals are inexpensive.;
- Normative biases may also shape interpretation, for instance, when analyses of DOA journals across citation-based quartiles emphasize the share in the highest quartile while ignoring overrepresentation in other (lower) categories.

A more balanced approach would view DOA as a highly diverse set of publishing practices (van Bellen & Céspedes 2025, p. 109) with both strengths and weaknesses, rather than presuming that it is inherently scholarly-led, non-commercial, and equitable. From such a starting point, research could investigate the diversity of the DOA landscape, the characteristics of different types of DOA journals, and the conditions under which particular models succeed, produce undesirable outcomes, or even fail.

A second need for emancipation relates to the funding logic of DOA studies. A lot of DOA research is oriented toward shaping scholarly publishing and thus required to deliver rapid, policy-relevant results. This corresponds with quantitative research designs using bibliometric, survey, and statistical methods, sometimes supplemented by qualitative elements that highlight singular aspects. As a consequence, several aspects of DOA journals that are more difficult and time-consuming to study remain underexplored. These would require research designs that allow for more contextualized perspectives. Four such directions are sketched below:

First, quantitative studies show that in many Western countries, DOA journals are concentrated in specific disciplines, most notably the social sciences and humanities (SSH), while in several Latin American countries the distribution differs. The reasons for this disciplinary focus remain unclear. In Norway (Frantsvåg 2022) and Canada (van Bellen & Céspedes 2025, p. 105), funding programs dedicated to SSH journals may partly explain this concentration. However, similar disciplinary patterns also occur in contexts without such programs, suggesting that country-level funding mechanisms (Zlodi et al. 2023, p. 21) is not the sole factor. Additional explanations may include the multiparadigmatic epistemic structure of SSH, the relatively small size of research communities (which makes individually initiated journals more viable), the importance of local languages, or the limited commercial interest of large publishers in these fields. Understanding these causes is essential for assessing whether DOA models can succeed in other disciplines.

Second, studies indicate that most DOA journals are small to mid-sized and that large-scale journals are virtually absent (see section 4.2). For a realistic assessment of the potential and limitations of DOA journals, it is necessary to investigate the factors influencing journal size and the mechanisms that may limit growth. Such questions require a comprehensive empirical design covering internal operations, cooperation among actors, disciplinary environments, and the funding and support structures from which necessary resources are drawn.

Third, as discussed in Section 9, several studies show that many DOA journals rely on what is described as 'voluntary' work and that many of them are not stable (e.g., Björk et al., 2016; Bosman et al., 2021). Both aspects, including their potential interrelations, require a more in-depth investigation of the internal organization of DOA journals and editorial offices. A productive starting point would be to move away from the notion of 'voluntary' work and instead focus on its unpaid character. This subtle shift enables questions about why scholars contribute without payment. Possible answers may include dependency relations within academic fields or institutions (Peacock 2016), path dependencies created by previous engagement (Taubert et al., 2024), and normative commitments. The work of Marcel Mauss (1954) on the ambivalence of gifts in general provides a fruitful conceptual lens for studying the effects of unpaid contributions. Empirical research could investigate ambivalent effects at the level of unpaid contributors and at the level of journal operation, shedding light on DOA journal instability. Such ambivalences may involve the retention of skilled personnel (Stojanovsky & Mofardin, 2025, p. 14), concerns about reliance on volunteers (Bosman et al., 2021, p. 124), workload transfers and burdening of highly committed team members (Taubert et al., 2024, p. 217), as well as instances of self-exploitation (Hahn et al., 2023, p. 47).

Fourth, scholarly publishing is a resource-intensive activity, and this also applies to DOA journals even when commercial publishers are absent and institutional publishers take the lead. Current cost estimations are not entirely convincing, as reported budgets and numbers of paid staff often do not align. Such inconsistencies raise questions about survey-based strategies that collect self-reported cost information from editors (Bosman et al., 2021; Hahn et al., 2022). This problem is particularly salient when practical tasks – "(such as running an online platform, generating visibility, typesetting etc.), labor that is outsourced to (paid) professionals by the commercial journals now needs to be done by academics themselves, on top of the unpaid labor they were already delivering" (van Charldorp & Labrie, 2024, p. 2). Realistic estimations of DOA costs must therefore account for unpaid labor, which generates opportunity costs for both individuals and institutions: time spent working for a journal is time not spent writing articles, developing proposals, completing PhD theses, or conducting research. Because of such opportunity costs, crucial components of DOA cost structures remain invisible in current data-gathering approaches. More ambitious research designs are needed – incorporating editors, editorial staff, support personnel, librarians, IT departments, and service providers—to collect information about budgeted and non-budgeted costs, unpaid labor, and in-kind contributions, and to cross-validate them.

## 11. Conclusions

This paper has reviewed the literature on DOA journals that was produced and published after this journal segment became a priority in European research policy around 2020. The majority of the literature consists of quantitative studies, examining the role of DOA journals through various aspects such as the number of journals, their publication output, the diversity of the landscape in terms of subject areas, languages, and publishing entities, indexing in major databases, awareness and perception among scholars, cost analyses, as well as insights into their internal operations. Without summarizing the results, two points deserve particular emphasis in this conclusion.

First, research on DOA journals is at least partly influenced by the science policy discourse surrounding them. The normativity inherent in this discourse sometimes spills over into research itself. In light of current frustrations with other pathways towards OA (APC-based OA and transformative agreements) together with the well-deserved recognition of contributors to DOA journals, the adoption of positive normative connotations by researchers is understandable. However, such an approach may hinder our understanding of these journals if it leads to biased interpretations and tendencies to neglecting their weaknesses. Furthermore, the temporality of DOA transformation requires research that is both timely and practically relevant. However, if such time constraints become dominant, important aspects – such as the internal operations of DOA journals and the structural consequences of reliance on unpaid labor – remain understudied.

Second, as shown in Section 6, the indexing of DOA journals in major bibliometric databases and OA evidence sources remains low. Consequently, exploratory studies of the DOA landscape often rely on alternative data sources, such as journal platforms (Beigel et al., 2024), journals using specific software (Khanna et al., 2022), or the database Dimensions (van Bellen & Cespedes, 2025). Such studies reveal that the proliferation of scientific publications from entire world regions is in large parts virtually eclipsed by the coverage of major databases. Scholarly communication on a global scale is therefore not only larger but also more diverse and fragmented along local languages and regional infrastructures. This is the crucial insight that DOA studies contribute to our broader understanding of the worldwide system of scholarly communication.

**Funding:** This research received no external funding.

**Conflicts of Interest:** The authors declare no conflicts of interest.

## Abbreviations

The following abbreviations are used in this manuscript:

| | |
|---|---|
| APC | Article Processing Charges |
| CRIS | Current Research Information Systems |
| DOA | Diamond Open Access |
| DOAJ | Directory of open access journals |
| DOAS | Diamond Open Access Standard |
| ERA | European Research Area |
| EZB | Electronische Zeitschriften Bibliothek Regensburg |
| FTE | Full Time Equivalent |
| IPSP | Institutional Publisher and Service Provider |
| JIF | Journal Impact Factor |
| OA | Open Access |
| OJS | Open Journal Systems |
| PKP | Public Knowledge Project |
| PMC | PubMed Central |
| ROAD | Directory of Open Access Scholarly Resources |
| WoS | Web of Science |